\documentclass[3p,times,procedia]{elsarticle}
\usepackage{aux/nupha_ecrc}
\volume{00}
\firstpage{1}
\journalname{Nuclear Physics A}
\runauth{}
\jid{nupha}
\jnltitlelogo{Nuclear Physics A}
\usepackage{amssymb,amsmath}

\usepackage[figuresright]{rotating}

\newcommand{\sig} 		{\sigma}
\newcommand{\kap} 		{\kappa}

\newcommand{\muB} 		{\mu_{_{\scriptstyle B}}}
\newcommand{\Nchi} 		[1]{\chi_{_{\!\scriptstyle{#1}}}}

\newcommand{\pdif} 	[1][]{\partial{#1}}              	% pd
\newcommand{\eV}   	{\;\mbox{e\hspace{-.12em}V}}     	% eV
\newcommand{\GeV}  	{\;\mbox{G\!\eV}}                % GeV
\newcommand{\Pt}  		{p_{_{\!{T}}}}
\begin{document}
%%%%%%%%%%%%%%%%%%%%%%%%%%%%%%%%%%%%%%%%%%%%%%%%%%%%%%
\begin{frontmatter}

 \title{Title\tnoteref{label1}}

%% Instructions from Editor: Please use the following \dochead only in the preprint version (e-print arXiv etc.); 
%% use empty \dochead{} when submitting to Nuclear Physics A!
\dochead{XXVIth International Conference on Ultrarelativistic Nucleus-Nucleus Collisions\\ (Quark Matter 2017)}
\dochead{}
%% e.g. \dochead{17th International Conference on Dynamical Processes in Excited States of Solids}

\title{Baryon susceptibilities from a holographic equation of state}


\author{Israel Portillo}
\address{Department of Physics, University of Houston, TX 77204, USA}

%%%%%%%%%%%%%%%%%%%%%%%%%%%%%%%%%%%%%%%%%%%%%%%%%%%%%%
\begin{abstract}
A black hole holographic model is used to mimic the behavior of the quark gluon plasma at finite temperature and density. This model reproduces lattice data at \(\muB\!=\!0\) and displays critical behavior at large densities. High order baryon susceptibilities are used to extract freeze-out points by comparing those susceptibilities with the corresponding net-protons distribution measured by STAR. Possible experimental signatures of the critical end point are discussed.
\end{abstract}
%%%%%%%%%%%%%%%%%%%%%%%%%%%%%%%%%%%%%%%%%%%%%%%%%%%%%%


\end{frontmatter}
% \linenumbers

%%%%%%%%%%%%%%%%%%%%%%%%%%%%%%%%%%%%%%%%%%%%%%%%%%%%%%
\section{Introduction}\label{intro}
%%%%%%%%%%%%%%%%%%%%%%%%%%%%%%%%%%%%%%%%%%%%%%%%%%%%%%
    
One of the primary goals of heavy ion collisions is the mapping of the phase diagram of quantum chromodynamics (QCD) at high temperatures ($T$) and baryonic chemical potential (\(\muB\)).  Lattice QCD has established that the transition from hadronic matter to the quark gluon plasma (QGP) at zero \(\muB\) is a crossover~\cite{Aoki:2006we}. With increasing $\muB$, the crossover is expected to end in a critical endpoint (CEP) where a first order phase transition begins.  The question of both the existence and the location of the CEP is fundamental to understanding QCD matter.

Experimentally, the Beam Energy Scan at RHIC is exploring unprecedented high density regions of the QCD phase diagram looking for experimental signatures of the CEP. Unfortunately, locating the CEP from first principles calculation is a formidable challenge due to the Fermi-sign problem. Nevertheless, higher-order baryonic susceptibilities from lattice QCD have been determined to reconstruct thermodynamic quantities at finite $\mu_B$ and aid searches for the CEP \cite{Allton:2002zi,Gavai:2008zr,Wu:2006su,D'Elia:2002gd,deForcrand:2002hgr,Gunther:2016vcp}. However,  truncation errors may make it very difficult to find the precise location of the critical point.

An alternative theoretical approach to explore the behavior of strongly interacting matter with a critical point is the holographic duality \cite{Maldacena:1997re}. This method was employed in Ref.\ \cite{Gubser:2008ny} to construct black hole solutions of higher dimensional gravitational theories with thermodynamic properties that mimic the QGP equilibrium properties computed on the lattice at $\muB\!=\!0$. The generalization of this type of model to include a baryonic charge was done in \cite{DeWolfe:2010he,DeWolfe:2011ts}, where it was shown that these holographic models can display a CEP at large baryon densities. These ``black hole engineered" non-conformal models possess a nonzero bulk viscosity \cite{Gubser:2008yx,Finazzo:2014cna}, which plays an important role in hydrodynamic simulations \cite{NoronhaHostler:2008ju,Noronha-Hostler:2013gga,Noronha-Hostler:2014dqa,Ryu:2015vwa}, and they can be used to compute baryonic susceptibilities and transport coefficients at nonzero $\muB$ \cite{Rougemont:2015ona,Rougemont:2015wca}. 

In a recent publication \cite{Finazzo:2016mhm}, the parameters of this black hole model were modified to provide a better match to current lattice data at $\muB\!=\!0$. The details of the corresponding extension to $\muB\!\neq\! 0$ will be published elsewhere \cite{newpaper}. In these proceedings (as was done in Ref.~\cite{Portillo:2016fso}), the baryonic susceptibilities from this refined black hole model is compared to the fluctuations of net-protons \cite{Adamczyk:2013dal} and  the corresponding $(T,\muB)$ freeze-out line is estimated. 

%%%%%%%%%%%%%%%%%%%%%%%%%%%%%%%%%%%%%%%%%%%%%%%%%%%%%%
\section{Baryon Number Susceptibilities}\label{intro}
%%%%%%%%%%%%%%%%%%%%%%%%%%%%%%%%%%%%%%%%%%%%%%%%%%%%%%
The baryon number susceptibility, \(\Nchi{n} = \Nchi{n}^B(T,\muB)\), can be numerically calculated in the black hole model at arbitrarily high \(\muB\). It is obtained by taking the \(n\) derivative of the pressure with respect to the baryonic chemical potential
\begin{align}
	 \Nchi{n} = \frac{\pdif^n}{\pdif(\muB/T)^n}\left(\frac{P}{T^4}\right)\,.
\end{align}
Susceptibilities are related directly to the moments of the distribution, from which it is convenient to form volume-independent susceptibility ratios 
\begin{align}
\begin{aligned}
    	&&\text{mean : }&     &	           M &\;=\; \Nchi{1} &&&&&&
    	& M/\sig^2 		 &\;=\; \Nchi{1}/\Nchi{2} &&\\
    	&&\text{variance : }&		& \sig^2 &\;=\; \Nchi{2} &&&&&&
    	& S\sig 			 	 &\;=\; \Nchi{3}/\Nchi{2}&&\\
    	&&\text{skewness : }&	&              S &\;=\; \Nchi{3}/\Nchi{2}^{3/2} &&&&&&
    	& \kap\sig^2  	 &\;=\; \Nchi{4}/\Nchi{2}&&\\
    	&&\text{kurtosis : }&		&     \kap  &\;=\; \Nchi{4}/\Nchi{2}^2 &&&&&&
    	& S\sig^3/M 		 &\;=\; \Nchi{3}/\Nchi{1}&&
\end{aligned}
\end{align}
The susceptibilities provide information about the effective degrees of freedom of a system and are essential to the characterization of phase transitions. Due to event-by-event fluctuations of the initial conditions in heavy-ion collisions, fluctuations of conserved charges occur on an event-by-event basis so a distribution is formed.  The moments of this distribution are fixed at the chemical freeze-out such that by comparing those moments with the susceptibilities one may extract \(T\) and \(\muB\) at freeze-out~\cite{Borsanyi:2014ewa,Alba:2014eba,Noronha-Hostler:2016rpd}. In this aspect, it is important to remark that the susceptibilities scale with different powers of the correlation length $\xi$, which diverges at the CEP. In fact, the higher order susceptibilities are more sensitive to the CEP and they diverge with higher powers of $\xi$. For instance, it was shown in Ref.\ \cite{Stephanov:2008qz} that for a homogeneous system in equilibrium,  $\Nchi{2}\!\sim\xi^2$, $\Nchi{3}\!\sim\! \xi^{9/2}$, and $\Nchi{4}\!\sim\xi^7$. In practice, the divergence of $\xi$ is limited in heavy ion collisions by the system size and finite time effects. Furthermore, it was argued in Ref.\ \cite{Stephanov:2011pb} that the ratio \(\kap\sig^2\) should show a non-monotonic behavior as one approaches the CEP. In the following section, this possibility is explored by analyzing the baryonic susceptibility ratios in our black hole model.

%%%%%%%%%%%%%%%%%%%%%%%%%%%%%%%%%%%%%%%%%%%%%%%%%%%%%%
\section{Results}\label{intro}
%%%%%%%%%%%%%%%%%%%%%%%%%%%%%%%%%%%%%%%%%%%%%%%%%%%%%%

\begin{figure}[h]
	\begin{center}
		\includegraphics[height=7.8pc]{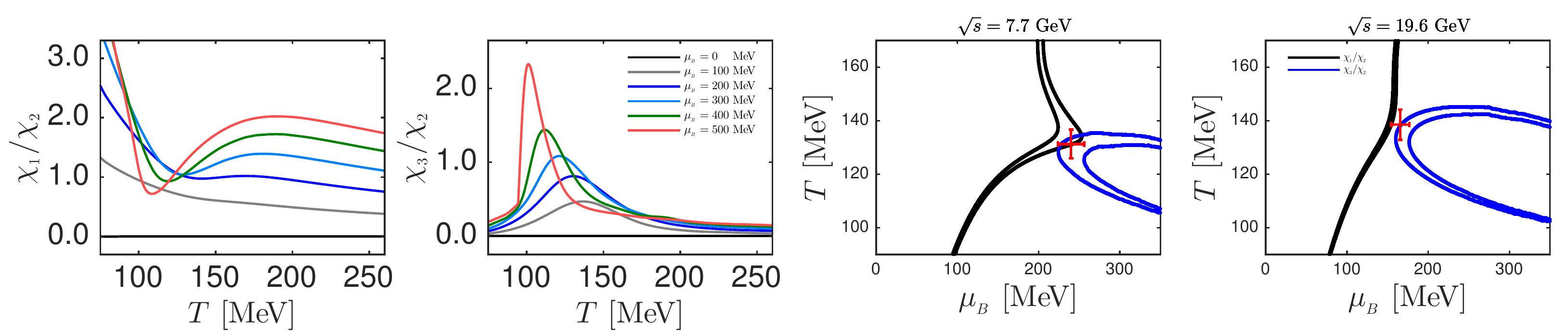}
	\end{center}
	\caption{\label{Fig:Ratios} (Color online) Susceptibility ratios, $\Nchi{1}/\Nchi{2}$ and $\Nchi{3}/\Nchi{2}$, obtained from the black hole model as a function of \(T\) for different values of \(\muB\) (left). Trajectories in the (\(T,\mu\)) plane that match the experimental values for the susceptibility ratios for a collision energy of $\sqrt{s}=7.7$\GeV and $\sqrt{s}=19.6$\GeV (left) obtained by STAR~\cite{Adamczyk:2013dal}, and the extracted freeze-out parameter (red cross).}
\end{figure}

The calculated black hole susceptibility ratios $\Nchi{1}/\Nchi{2}$ and $\Nchi{3}/\Nchi{2}$ are shown as function of \(T\) and different values of \(\muB\) in Fig.~\ref{Fig:Ratios} (left). From these ratios, bands in the \(T\) and \(\muB\) plane were obtained by imposing that they reproduce the corresponding experimental value from STAR (for energies \(\sqrt{s}=[7.7-200]\)\GeV). Fig.~\ref{Fig:Ratios} (right) also shows those trajectories for a collision energy of $\sqrt{s}=7.7$\GeV\ and $\sqrt{s}=19.6$\GeV. One can see that some energies show a clear overlap while others do not, which may be attributed to other effects such as decays or acceptance cuts not included in our model. In both cases, a freeze-out parameter (red cross) is determined by the area of closest distance between the farthest boundaries of the curves.

\begin{figure}[h]
	\begin{center}
		\includegraphics[height=10pc]{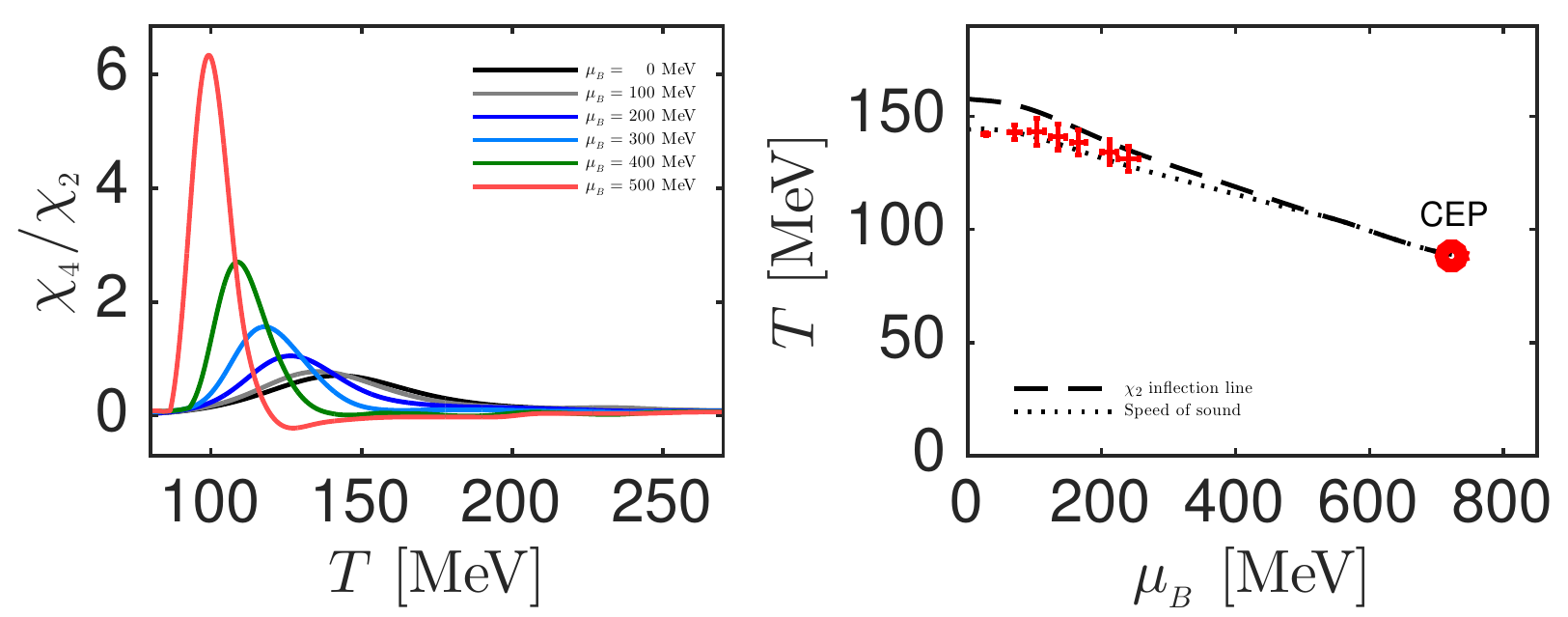}
		\caption{\label{Fig:RatPhDia}(Color online) Susceptibility ratios $\Nchi{4}/\Nchi{2}$ obtained from the black hole model (left panel) at finite density. Freeze-out parameters (red data points) extracted by comparing the model calculations to STAR data \cite{Adamczyk:2013dal}; the location of the minimum of $c_s^2$ (dashed line) and the inflection point of $\Nchi{2}$ (solid line) of the black hole model~\cite{newpaper}, across $\muB$, are also shown (left panel). A preliminary estimate for the location of the CEP of the black hole model is also shown.}
	\end{center}
\end{figure}

Fig.\ \ref{Fig:RatPhDia} shows the susceptibility ratio $\Nchi{4}/\Nchi{2}$ (left) as a function of \(T\) for different values of \(\muB\). This ratio diverges faster than $\Nchi{2}$ since \(\Nchi{4}\) has a stronger dependence on the correlation length as ones approaches the CEP at larger baryon chemical potential. For \(\muB\!=\!500\) MeV its peak is already three times larger than in the ratio $\Nchi{3}/\Nchi{2}$. Those curves also show that as \(\muB\) approaches the large values close to the CEP of the model a kink is developed. This behavior is consistent with the  non-monotonicity expected from Ref.\ \cite{Stephanov:2011pb}. 

Fig.~\ref{Fig:RatPhDia} (right) shows the phase diagram obtained in the black hole model. The estimate for the CEP is shown, as well as the extracted freeze-out parameters, which are placed between the inflection point of \(\Nchi{2}\)(dashed line) and the minimum of the speed of sound squared $c_{s}^2$ (dotted line). The freeze-out chemical potentials found at the collision energy of \(\sqrt{s}=19.6\)\GeV\ and lower are smaller that the ones traditionally quoted from the thermal fits. This can be attributed to the fact that the holographic approach does not have strangeness and electric charge chemical potentials, which have larger effect at high densities. Besides, our model does not include any acceptance cuts, and it compares baryon number fluctuations to the experimental data from net-protons.

\begin{figure}[h]
	\begin{center}
		\includegraphics[height=10pc]{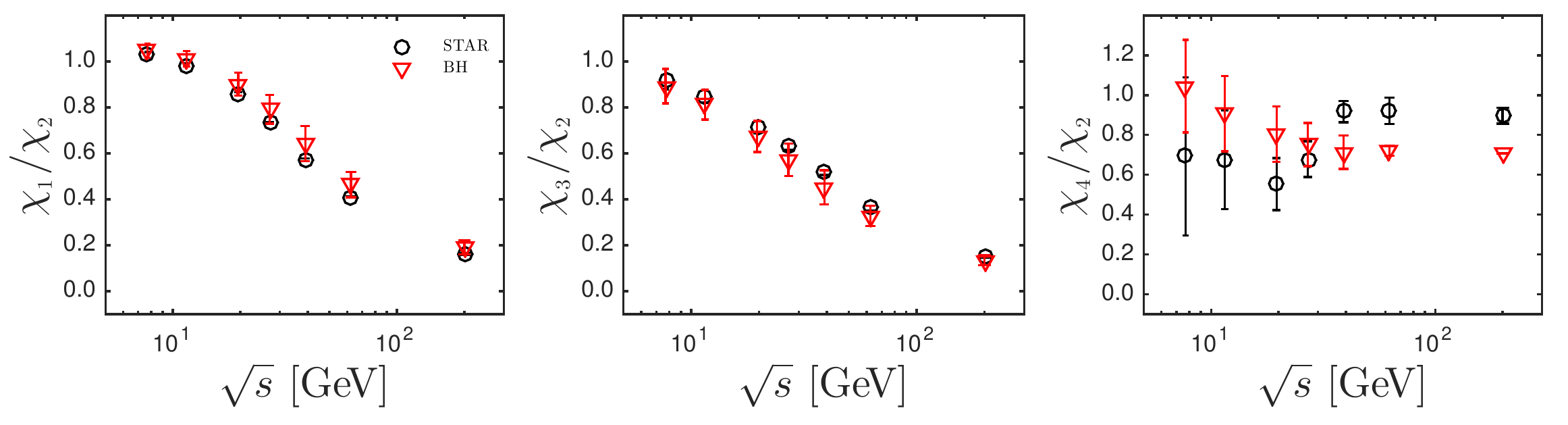}
	\end{center}
    \caption{\label{Fig:Pred}(Color online) Susceptibility ratios calculated along the freeze-out line in the black hole model (red arrows) as a function of collision energy \(\sqrt{s}\) compared with the corresponding net-proton distribution form STAR~\cite{Luo:2015ewa} (black circles).}
\end{figure}

In Fig.\ \ref{Fig:Pred} our model calculations, computed at freeze-out, for \(\Nchi{1}/\Nchi{2}$, $\Nchi{3}/\Nchi{2}$, and $\Nchi{4}/\Nchi{2}$ (red arrows)  are shown in comparison  with the corresponding experimental data points from STAR \cite{Adamczyk:2013dal} (black circles).
It is important to point out here that only the first two ratios were used to extract the freeze-out parameter. Thus, the value of the ratio $\Nchi{4}/\Nchi{2}$ at freeze-out is a prediction of our model. This ratio follows the trend of the experimental data at low $\sqrt{s}$ and also shows a hint of non-monotonic behavior at higher $\sqrt{s}$-values compared with the non-monotonicity of the experimental data.

%%%%%%%%%%%%%%%%%%%%%%%%%%%%%%%%%%%%%%%%%%%%%%%%%%%%%%
\section{Conclusions}\label{conc}
%%%%%%%%%%%%%%%%%%%%%%%%%%%%%%%%%%%%%%%%%%%%%%%%%%%%%%

We used preliminary data from a new holographic model \cite{newpaper} to compute high order susceptibilities and extract the freeze-out parameters across collision energies. The computed $\Nchi{4}/\Nchi{2}$ ratio along the freeze-out line shows a hint of non-monotonic behavior before it grows with decreasing $\sqrt{s}$. This behavior appears even though the freeze-out values are still far from the high density region where the CEP of the model is located, as it will be shown in \cite{newpaper}. Preliminary STAR data with higher $\Pt$ cuts ($0.4<\Pt<2.0$ GeV) \cite{Luo:2015ewa} show a stronger enhancement for $\Nchi{4}/\Nchi{2}$ at low $\sqrt{s}$ and a more pronounced non-monotonicity. In a future study, we plan to examine this new data taking into account the acceptance cuts and decays. More details about the holographic model and its applications to the study of the QCD critical point, together with comparisons to experimental data, will be given in \cite{newpaper}.

\section{Acknowledgements}

I would like to thank the organizers of QM2017 for selecting my poster for a flash talk and, especially,  my collaborators on this work: R.\ Critelli, J.\ Noronha, J.\ Noronha-Hostler, C.\ Ratti, and R.\ Rougemont.  
This material is based upon work supported by the National Science Foundation under grant AC02-06CH11357.  I gratefully acknowledge the use of the Maxwell Cluster and the advanced support from the Center of Advanced Computing and Data Systems at the University of Houston.

\bibliographystyle{elsarticle-num}
\bibliography{Isra_QM17.bib}

\end{document}